\documentclass[MSNbibl,nameyear,seceqn,dvips]{arxstspdf}
\usepackage{dcolumn}
\usepackage{graphicx}
\usepackage{flushend}
\usepackage{stfloats}
\volume{28}
\issue{1}
\pubyear{2013}
\firstpage{95}
\lastpage{115}
\doi{10.1214/12-STS403} 

\makeatletter
\newcolumntype{d}[1]{D{.}{.}{#1}}

\newcommand{\rright}{\right}
\newcommand{\lleft}{\left}
\setattribute{ead}{sep}{; }
\renewcommand{\citep}[1]{(\citeauthor{#1}, \citeyear{#1})}
\newtheorem{theorem}{Theorem}[section]
\newproclaim{remark}{Remark}
\newproclaim{fact}{Fact}
\makeatother

\begin{document}
\begin{frontmatter}

\title{Bayes Model Selection with Path Sampling: Factor Models and
Other Examples}
\runtitle{Bayes Model Selection}

\begin{aug}
\author[a]{\fnms{Ritabrata} \snm{Dutta}\corref{}\ead[label=e1]{rdutta@purdue.edu}\ead[label=e11]{statrita2004@gmail.com}}
\and
\author[a]{\fnms{Jayanta K.} \snm{Ghosh}\ead[label=e2]{ghosh@purdue.edu}}
\runauthor{R. Dutta and J. K. Ghosh}

\affiliation{Purdue University}

\address[a]{Ritabrata Dutta is Ph.D. Student and
Jayanta K. Ghosh is Professor, Department of Statistics, Purdue University,
Lafayette, Indiana 47907, USA \printead{e1,e11,e2}.}

\end{aug}

%
\begin{abstract}
We prove a theorem justifying the regularity conditions which are
needed for Path Sampling in Factor Models. We then show that the
remaining ingredient, namely, MCMC for calculating the integrand at
each point in the path, may be seriously flawed, leading to wrong
estimates of Bayes factors. We provide a new method of Path Sampling
(with Small Change) that works much better than standard Path Sampling
in the sense of estimating the Bayes factor better and choosing the
correct model more often. When the more complex factor model is true,
PS-SC is substantially more accurate. New MCMC diagnostics is provided
for these problems in support of our conclusions and recommendations.
Some of our ideas for diagnostics and improvement in computation
through small changes should apply to other methods of computation of
the Bayes factor for model selection.
\end{abstract}

%
\begin{keyword}
\kwd{Bayes model selection}
\kwd{covariance models}
\kwd{path sampling}
\kwd{Laplace approximation}
\end{keyword}

\end{frontmatter}

\section{Bayes Model Selection}\label{sec1}
Advances in MCMC techniques to compute the posterior for many complex,
hierarchical models have been a major reason for success in Bayes
modeling and analysis of complex phenomena (Andrieu, Doucet and Robert, \citeyear{r3}). These techniques
along with applications are surveyed in numerous papers, including \citet
{r7}, \citet{r26} and \citet{r33}. Moreover, many\break Bayesian books on
applications or theory and methods provide a quick introduction to
MCMC, such as \citet{r15}, \citet{r19}, \citet{r12} and \citet{r28}.

Just as the posterior for the parameters of a given model are
important for calculating Bayes estimates, posterior variance,
credibility intervals and a general description of the uncertainty
involved, one needs to calculate Bayes factors for selecting one of
several models. Bayes factors are the ratio of marginals of given data
under different models, when more than one model is involved and one
wishes to choose one from among them, based on their relative or
posterior probability. The ratio of marginals measures the relative
posterior probability or credibility of one model with respect to the
other if we make the usual objective choice of half as prior
probability for each model.

Although there are many methods for calculating Bayes factors, their
success in handling complex modern models is far more limited than
seems to be generally recognized. Part of the reason for lack of
awareness of this is that model selection has become important
relatively recently. Also, one may think that, in principle,
calculation of a BF can be reduced to the calculation of a posterior,
and hence solvable by the same methods as those for calculating the
posterior. Reversible Jump MCMC (RJMCMC) is an innovative methodology
due to \citet{r20}, based on this simple fact. However, two models
essentially lead to two different sets of states for any Markov chain
that connects them. The state spaces for different models often differ
widely in their dimension. This may prevent good mixing and may show up
in the known difficulties of tuning RJMCMC. For a discussion of tuning
difficulties see \citet{r33}.

Another popular method for calculating BF is path sampling (PS), which
is due to \citet{r14} and recently re-examined by \citet{r24}. Our major
goal is to explore PS further in the context of nested, relatively
high-dimensional covariance models, rather than nonnested
low-dimen\-sional mean models, as in the last reference. The new examples
show both similarities and sharp changes from the sort of behavior
documented in \citet{r24}.

We consider three paths, namely, the geometric mean path, the
arithmetic mean path and the parametric arithmetic mean path, which
appear in \citet{r14}, \citet{r24}, \citet{r17}, \citet{r18}, \citet{r23}
and \citet{r36}. Other applications of path sampling and bridge sampling
(with some modifications) appear in \citet{r22b}, \citet{r11b}, \citet
{r36b} and \citet{r36c}. Our priors are usually the diffuse Cauchy
priors, first suggested by \citet{r22} and since then recommended by
many others, including Berger (personal communication), \citet{r25},
\citet{r16} and \citet{r18}. But we also examine other less diffuse
priors too, going all the way to normal priors. Since \citet{r24} have
studied applications of PS to mean like parameters, we focus on
covariance models. We restrict ourselves generally to factor models for
covariance, which have become quite popular in recent applications, for
example, \citet{r27}, \citet{r17}, \citet{r18} and \citet{r23}. The recent
popularity of factor models is due to the relative ease with which they
may be used to provide a sparse representation of the covariance matrix
of multivariate normal data in many applied problems of finance,
psychometry and epidemiology; see, for example, the last three
references. Also, often it leads to interesting scientific insight; see
\citet{r4}.

In addition to prior, likelihood and path, there are other choices to
be made before PS can be implemented, namely, a method of discretizing
the path, for example, by equispaced points or adaptively\break \citep{r24}
and how to integrate the score function of \citet{r14} at each point in
the discrete path. A popular method is to use MCMC. These more
technical choices are discussed later in the paper. Along with PS, we
will consider other methods like Importance Sampling (IS) and its
descendants like Annealed Importance Sampling (AIS), due to \citet{r30},
and Bridge Sampling (BS), due to \citet{r29}.

We now summarize our contribution in this paper.

In Section~\ref{sec2} we review what is known about path sampling and factor
models. We introduce factor models, a suitable path and suitable
diffuse $t$-priors. The path we use was first introduced in \citet{r14}
for mean models and by \citet{r23} and \citet{r18} for factor models.

In Section~\ref{sec2.4} we prove a theorem (Theorem~\ref{th2.1}) which essentially
shows that except for the convergence of MCMC estimates for expected
score function $E_t(U(\theta,t))$ at each grid point $t$ in the path, all
other needed conditions for PS will hold for our chosen path, prior and
likelihood for factor models. In one of the remarks following the
theorem we generalize this result to other paths. Remark~\ref{rem3} points to
the need for some finite moments for the prior, not just for Theorem
\ref{th2.1} to hold but for the posterior to behave well. Then in Remark~\ref{rem5} we
provide a detailed, heuristic argument as to why the MCMC may fail
dramatically by not mixing properly if the data has come from the
bigger of the two models under consideration. If our heuristics is
correct, and there is a small interval where $E_t(U(\theta,t))$
oscillates most, then a grid size that is a bit coarse will not only be
a bit inaccurate, it will be very wrong. Even if the grid size is
sufficiently small, one will need to do MCMC several times with
different starting points just to realize PS will not work. Our new
proposal avoids these problems but will require more time if many
models are involved.

In Section~\ref{sec3} we give an argument as to why the above is unlikely to be
true if the data has come from the smaller model. More importantly, in
Section~\ref{sec3.3} we propose a modification of PS, which we call Path
Sampling with Small Change (PS-SC) which is expected to do better.

Implementation of PS and PS-SC can be very time consuming due to the
need of MCMC sampling for each grid point along the path. Time can be
saved if we can implement PS and PS-SC by parallel computation, as
noted by \citet{r14}.

In Section~\ref{sec3.4} we show MCMC output for the various cases discussed
and validate our heuristics above. The diagnostics via projection into
likelihood space should prove useful for other model selection
problems. Our gold standard is PS-SC, based on an MCMC with the number
of draws $m=50\mbox{,}000$ and burn-in of 1000, if necessary. But actually in
our examples $m=6000$ and burn-in of 1000 suffices for PS-SC. For
other model selection rules we also go up to $m=50\mbox{,}000$ if necessary.
After Section~\ref{sec3.4}, having shown our modified PS, namely, PS-SC, is
superior to PS under both models, we do not consider PS in the rest of
the paper.

In the last two sections we touch on the following related topics:
effects of grid size, alternative path, alternative methods and
performance of PS-SC and some other methods in very high-dimensional
simulated and real examples. PS-SC seems to choose the true models in
the simulated cases and relatively conservative models for real data.
In Section~\ref{sec5} we explore various real life and high-dimensional factor
models, with the object of combining PS-SC with two of the methods
which do relatively well in Section~\ref{sec4} to reduce the time of PS-SC in
problems with the number of factors rather high, say, 20 or 26, for
which PS-SC can be quite slow. For these high-dimensional examples, we
use Laplace approximation to marginals for preliminary screening of
models. A few general comments on Laplace approximation in
high-dimensional problems are in Section~\ref{sec5}.

In Appendix~\ref{appa1} we introduce briefly a few other methods like Annealed
Importance Sampling (AIS) which we have compared with PS-SC. Finally,
Appendix~\ref{appa2} points to some striking differences between what we
observe in factor models and what one might have expected from our
familiar classical asymptotics for maximum likelihood estimates. Of
course, as pointed out by \citet{r11}, classical asymptotics does not
apply here, but it surprised us that the differences would be so stark.
It is interesting and worth pointing out that the Bayes methods like
PS-SC can be validated partly theoretically and partly numerically in
spite of a lack of suitable asymptotic theory.

\section{Path Sampling and Factor Models}\label{sec2}
In the following subsections we review some basic facts about PS,
including the definition of the three paths and the notion of an
optimal path. More importantly, since our interest would be in model
selection for covariance rather than mean, we introduce factor models
and then PS for factor models in Sections~\ref{sec2.3} and~\ref{sec2.4}.

Section~\ref{sec2.1} is mostly an introduction to PS and reviews previous
work. After that we show the failure of PS-estimates in a toy problem
related to the modeling of the covarince matrix in Section~\ref{sec2.2}. In
Section~\ref{sec2.3} we introduce factor models and our priors. Section
\ref{sec2.4} introduces paths that we consider for factor models and a theorem
showing the regularity conditions needed for validity of PS under
factor models. Then in a series of remarks we extend the theorem and
also study and explain how the remaining ingredient of PS, namely,
MCMC, can go wrong. We show a few MCMC outputs to support our arguments
in Section~\ref{sec3.4}. This particular theme is very important and will
come up several times in later sections or subsections where related
different aspects will be presented.
\subsection{Path Sampling}\label{sec2.1}
Among the many different methods related to importance sampling, the
most popular is Path Sampling (PS). However, PS is best understood as a
limit of the simpler Bridge Sampling (BS) (Gelman and Meng, \citeyear{r14}). So we first
begin with BS.

It is well known that unless the densities of the sampling and target
distributions are close in relative importance sampling weights,
Importance Sampling (IS) will have high variance as well as high bias.
Due to the difficulty of finding a suitable sampling distribution for
IS, one may try to reduce the difficulty by introducing a
nonnormalized intermediate density $f_{1/2}$ that acts like a bridge
between the nonnormalized sampling density $f_1$ and nonnormalized
target density $f_0$ \citep{r29}. One can then use the identity
$Z_1/Z_0=\frac{Z_{1/2}/Z_0}{Z_{1/2}/Z_1}$ and estimate both the
numerator and denominator by IS. Extending this idea, \citet{r14}
constructed a whole path $f_t\dvtx t\in[0,1]$ connecting $f_0$ and $f_1$.
This is also like a bridge. Discretizing this, they get the identity
$Z_1/Z_0=\prod^L_{l=1}{\frac
{Z_{(l-1/2)}/Z_{(l-1)}}{Z_{(l-1/2)}/Z_{(l)}}}$, which leads to a chain
of IS estimates in the numerator and denominator. We call this estimate
the Generalized Bridge Sampling (GBS) estimate.

More importantly, \citet{r14} introduced PS, which is a new scheme,
using the idea of a continuous version of GBS but using the log scale.
The PS estimate is calculated by first constructing a path as in BS.
Suppose the path is given by $p_t\dvtx  t\in[0,1]$ where for each $t$,
$p_t$ is a probability density. Then we have the following definition:
%
\begin{equation}
\label{eqn21} p_t(\theta)=\frac{1}{z_t}f_t(\theta),
\end{equation}
where $f_t$ is an unnormalized density and $z_t= \break\int{f_t(\theta)\,d\theta
}$ is the normalizing constant. Taking the derivative of the logarithm
on both sides, we obtain the following identity under the assumption of
interchangeability of the order of integration and differentiation:
%
\begin{eqnarray}
\label{eqn22}\qquad \frac{d}{dt}\log(z_t)&=&\int{\frac{1}{z_t}
\frac{d}{dt}f_t(\theta)\mu (d\theta)}
\nonumber
\\[-8pt]
\\[-8pt]
\nonumber
&=&E_t\biggl[
\frac{d}{dt}\log{f_t(\theta)}\biggr]=E_t\bigl[U(
\theta,t)\bigr],
\end{eqnarray}
where the expectation $E_t$ is taken with respect to $p_t(\theta)$ and
$U(\theta,t)=\frac{d}{dt}\log{f_t(\theta)}$. Now integrating (\ref
{eqn22}) from~0 to 1 gives the log of the ratio of the normalizing
constants, that is, log BF in the context of model selection:
%
\begin{equation}
\label{eqn23} \log\biggl[\frac{Z_1}{Z_0}\biggr]=\int_0^1
E_t\bigl[U(\theta,t)\bigr]\,dt.
\end{equation}
To approximate the integral, we discretize the path with~$k$ points
$t_{(0)}=0<t_{(1)}<\cdots<t_{(k)}=1$ and\break draw~\textit{m} MCMC samples
converging to $p_t(\theta)$ at each of these $k$ points. Then estimate
$E_t[U(\theta,t)]$ by\break $\frac{1}{m}\sum{U(\theta^{(i)},t)}$ where $\theta^{(i)}$
is the MCMC output. To estimate the final log Bayes factor,
commonly numerical integration schemes are used. It is clear that MCMC
at different points ``$t$'' on the path can be done in parallel. We
have used this both for PS and for our modification of it, namely,
PS-SC introduced in Section~\ref{sec3.3}.

\citet{r14} showed there is an optimum path in the whole distribution
space providing a lower bound for MCMC variance, namely,
\begin{eqnarray*}
\biggl[\arctan{\frac{H(f_0,f_1)}{\sqrt{4-H^2(f_0,f_1)}}}\biggr]^2\Big/m,
\end{eqnarray*}
where $f_0$ and $f_1$ are the densities corresponding to the two models
compared and $H(f_0,f_1)$ is their Hel\-linger distance. Unfortunately in
nested examples $f_0$ and $f_1$ are mutually orthogonal, so
$H(f_0,f_1)$ takes the trivial value of two. Moreover, $m$ is so large
that the lower bound becomes trivial and unattainable. However, in a
given problem, one path may be more suitable or convenient than another.

Following \citet{r14} and \citet{r24}, we consider three paths generally
used for the implementation of PS. The Geometric Mean Path (GMP) and
Arithmetic Mean Path (AMP) are defined by the mean
[$f_t=f_0^{(1-t)}f_1^t$ and $f_t=tf_0+(1-t)f_1$, resp.] of the
densities of two competing models for each model $M_t\dvtx t\in(0,1)$ along
the path. Our notation for the Bayes factor is given later in equation (\ref{eq2.6}).

One more common path is obtained by assuming a specific functional form
$f_{\theta}$ for the density and then constructing the path in the
parametric space ($\theta\in\Theta$) of the assumed density. If $\theta_t=t\theta_0+(1-t)\theta_1$,
then $f_{t,\theta_t}$ is the density of
the model $M_t$, where $f_{0,\theta_0}=f_0$ and $f_{1,\theta_1}=f_1$.
We denote this third path as the Parametric Arithmetic Mean Path
(PAMP). The PAMP path was shown by \citet{r14} to minimize the Rao
distance in a path for model selection about normal means. More
importantly, it is very convenient for use of MCMC, as shown for some
factor models by \citet{r36} and \citet{r18}, and for linear models by
\citet{r24}. Implementation of PS with the paths mentioned above is
denoted as GMP-PS, AMP-PS and PAMP-PS. In view of the discussion in
\citet{r24} regarding the degeneracy of the AMP-PS, we will only
consider PAMP-PS and GMP-PS.

Unlike \citet{r24}, who study models about means, our interest is in
studying model selection for covariance models, specifically factor
models with different number of factors. These are discussed in the
Sections~\ref{sec2.3} and~\ref{sec2.4}. Performance of PS for covariance models can be
very different from the examples in \citet{r24}. In the next subsection
we give a toy example of covariance model selection where PS fails and
our proposed modification PS-SC is also not applicable.
\subsection{Covariance Model: Toy Example}\label{sec2.2}

To illustrate the difficulties in calculation of the BF that we discuss
later, we begin by considering a problem where we can calculate the
true value of the Bayes factor.

Assuming $Y_p\sim N(0,\Sigma)$, for some $m<p$ we wish to test whether
$Y_{1,\ldots,m}$ and $Y_{m+1,\ldots,p}$ are independent or not. If
$\Sigma=
\bigl({
{A_{11}\enskip A_{12}} \atop
{A'_{12} \enskip A_{22}} }\bigr)
$
where $Y_{1,\ldots,m}\sim N(0,A_{11})$ and $Y_{m+1,\ldots,p} \sim
N(0,A_{22})$, then the competitive models for a fixed \textit{m} will
be $M_0\dvtx  A_{12}=0$ vs $M_1\dvtx  A_{12}\neq0$. Under $M_1$ we use an
inverse-Wishart prior for the covariance matrix, as it helps us to
calculate the true BF, using the conjugacy property of the prior. Under
$M_0$ we take $A_{11}$, $A_{22}$ to be independent, each with an
inverse Wishart prior.

We illustrate the above problem with $p=10$ and $m=7$ for a positive
definite matrix $\Sigma^0=\bigl( {
{A^0_{11} \enskip A^0_{12}} \atop
(A^0_{12})' \enskip A^0_{22}
}\bigr)$
(given in Appendix~\ref{appa3}). We implement the path sampling for this
problem connecting $M_0$ and $M_1$, using a Parametric Arithmetic Mean Path:
%
\begin{eqnarray}\quad
M_t\dvtx y_i\sim N \biggl(0,\Sigma= \pmatrix{
A^0_{11} & tA^0_{12} \vspace*{2pt}
\cr
t\bigl(A^0_{12}\bigr)' & A^0_{22}
} \biggr).
\end{eqnarray}
For every $0\leq t \leq1$, the $\Sigma$ matrix is positive definite,
being a convex combination of two positive definite matrices. For $t=0$
and $t=1$ we get the models $M_0$ and $M_1$.

We can estimate the Bayes factor by using the path sampling schemes as
described earlier. We simulated two data sets, one each from $M_0$ and
$M_1$, and report the true BF value with the PS estimate in Table~\ref{tab1}.
Here the reported Bayes factor is defined as the ratio\vspace*{1pt} $\frac
{m_1}{m_0}$, where $m_1$ and $m_0$ are the marginals under the models
$M_1$ and $M_0$, respectively.

\begin{table}
\caption{Performance of PS in toy example modeling covariance: Log
Bayes factor (MCMC-standard deviation)}\label{tab1}
\begin{tabular*}{\columnwidth}{@{\extracolsep{\fill}}lcc@{}}
\hline
\textbf{Method} & \textbf{Data 1} & \multicolumn{1}{c@{}}{\textbf{Data 2}} \\
\hline
True BF value& 258.38& $-132.87$ \\
PS estimate of BF& 184.59 (0.012)& $-20.11$ (0.008)\\
\hline
\end{tabular*}
\end{table}

The values in the table show us that the estimated BF value is off by
an order of magnitude when $M_0$ is true. The value is relatively
stable as judged by the MCMC-standard deviation based on 10 runs and
near to the true value for $M_1$.

\subsection{Factor Models and Bayesian Specification of Prior}\label{sec2.3}

A factor model with $k$ factors is defined as $n$ i.i.d. observed r.v.'s
\[
y_i=\Lambda\eta_i+\varepsilon_i,\quad
\varepsilon_i \sim N_p(0,\Sigma),
\]
where $\Lambda$ is a $p\times k$ matrix of factor loadings,
\[
\eta_i=(\eta_{i1},\ldots,\eta_{ik})'
\sim N_k(0,I_k)
\]
is a vector of standard normal latent factors, and $\varepsilon_i$ is the
residual with diagonal covariance matrix $\Sigma=\operatorname{diag}(\sigma_1^2,\ldots,
\sigma_p^2)$. Thus, we may write the marginal distribution of $y_i$ as
$N_p(0,\Omega)$, $\Omega=\Lambda\Lambda'+\Sigma$. This model implies
that the sharing of common latent factors explains the dependence in
the outcomes and the outcome variables are uncorrelated given the
latent factors.

A factor model, without any other constraints, is nonidentifiable
under orthogonal rotation. Post-multiplying $\Lambda$ by an orthogonal
matrix $P$, where $P$ is such that $PP'=I_k$, we obtain exactly the same
$\Omega$ as in the previous factor model. To avoid this, it is
customary to assume that $\Lambda$ has a full-rank lower triangular
structure, restricting the number of free parameters in $\Lambda$ and
$\Sigma$ to $q=p(k+1)-k(k-1)/2$, where $k$ must be chosen so that
$q\leq p(p+1)/2$. The reciprocal of diagonal entries of $\Sigma$ forms
the precision vector here.

It is well known that maximum likelihood estimates for parameters in
factor models may lie on boundaries and, hence, likelihood equations
may not hold \citep{r2}. The Bayes estimate of $\Omega$ defined as
average over MCMC outputs is well defined, easy to calculate and, being
average of positive definite matrices, is easily seen to be positive
definite. This fact is used to search for maximum likelihood estimates
(mle) or maximum prior$\times$likelihood estimates (mple) in a
neighborhood of the Bayes estimate.

We also note for later use the following well-known simple fact, for
example, \citet{r2}. If the likelihood is maximized over all positive
definite matrices $\Omega$, not just over factor models, then the
global maximum for $n$ independent observations exists and is given by
%
\begin{equation}
\label{eqn24} \hat{\Omega}=\frac{1}{n-1}\sum^n_{i=1}(y_i-
\bar{y}) (y_i-\bar{y})'.
\end{equation}

From the Bayes model selection perspective, a specification of the
prior distribution for the free elements of $\Lambda$ and $\Sigma$ is
needed. Truncated normal priors for the diagonal elements of $\Lambda$,
normal priors for the lower triangular elements and inverse-gamma
priors for $\sigma^2_1,\ldots,\sigma_p^2$ have been commonly used in
practice due to conjugacy and the resulting simplification in posterior
distribution. Prior elicitation is not common.

\citet{r18} addressed the above problems by using the idea of \citet{r16}
to introduce a new class of default priors for the factor loadings that
have good mixing properties. They used the Gibbs sampling scheme and
showed there was good mixing and convergence. They used parameter
expansion to induce a class of $t$ or folded $t$-priors depending on sign
constraints on the loadings. Suitable $t$-priors have been very popular.
We use the same family of priors but consider a whole range of many
degrees of freedom going all the way to the normal and use the same
Gibbs sampler as in \citet{r17}. We have used a modified version of
their code.

In the factor model framework, we stick to the convention of denoting
the Bayes factor for two models with latent factors $h-1$ and $h$ as
%
\begin{equation}\label{eq2.6}
\mathit{BF}_{h, h-1}=\frac{m_{h}(x)}{m_{h-1}(x)},
\end{equation}
where ${m_{h}(x)}$ is the marginal under the model having $h$ latent
factors. \textit{So the Bayes factor for the simpler model \textup{(}defined as
$M_0$\textup{)} and complex model \textup{(}defined as $M_1$\textup{)} with $h-1$ and $h$ latent
factors will be defined as $\mathit{BF}_{h,h-1}$}. We choose the model with $h$
and $h-1$ latent factors, respectively, depending on the value of the
log Bayes factor being positive and negative. Alternatively, one may
choose a model only when the value of $\log{\mathit{BF}}$ is decisively negative
or positive, say, less than or greater than a chosen threshold.

\subsection{Path Sampling for Factor Models}\label{sec2.4}
There are several variants of path sampling which have been explored
in different setups, depending on choice of path, prior and other
tuning parameters (grid size and MCMC sample size). In the factor model
setup the parametric arithmetic mean path (PAMP) [used by \citet{r36}
and \citet{r18}] seems to be the most popular one. We also consider
Geometric Mean Path (GMP) along with the PAMP for the factor model.

By constructing a GM path from corresponding prior to the posterior,
we can estimate the value of the log-marginal under both $M_0$ and
$M_1$, which in turn leads us to an estimate of the log-BF. We will
first describe the two paths and their corresponding score functions to
be estimated along the path.

\begin{longlist}[(ii)]
\item[(i)] \textit{Parametric arithmetic mean path}: Lee and\break Song (\citeyear{r23}) used
this path in factor models, following an example in \citet{r14}. \citet
{r17} also used this path along with parameter expansion. Here we
define $M_0$ and $M_1$ to be the two models corresponding to the factor
model with factors $h-1$ and $h$, respectively, and then connect them
by the path $ M_t\dvtx  y_i=\Lambda_t\eta_i+\varepsilon_i, \Lambda_t=(\lambda_1,\lambda_2,\ldots,\lambda_{h-1},t\lambda_h)$,
where\vadjust{\goodbreak} $\lambda_j$ is the $j$th column of the loading matrix. So for
$t=0$ and $t=1$ we get the models $M_0$ and $M_1$. The likelihood
function at grid point $t$ is a MVN which is denoted as $f(Y|\Lambda
,\Sigma,\eta,t)$. We have independent priors $\pi(\Lambda), \pi(\Sigma
), \pi(\eta)$ and a score function,
%
\begin{eqnarray}
\label{eqn25} &&U(\Lambda, \Sigma, \eta,Y, t)
\nonumber
\\[-8pt]
\\[-8pt]
\nonumber
&&\quad=\sum^n_{i=1}{(y_i-
\Lambda_t\eta_i)'\Sigma^{-1}
\bigl(0^{p\times(h-1)},\lambda_h\bigr)\eta_i}.
\end{eqnarray}
For fixed and ordered grid points along the path
$t_{(0)}=0<t_{(1)}<\cdots<t_{(S)}<t_{(S+1)}=1$, our path sampling
estimate for the log Bayes factor is
%
\begin{eqnarray}
\label{26}&& \log(\widehat{\mathit{BF}}_{h:h-1})
\nonumber
\\[-8pt]
\\[-8pt]
\nonumber
&&\quad=\frac{1}{2}\sum
^{S}_{s=0}(t_{s+1}-t_{s}) \bigl(
\widehat{E}_{s+1}(U)+\widehat{E}_{s}(U)\bigr).
\end{eqnarray}
We simulate $m$ samples of $(\Lambda_{t_s,i},\Sigma_i,\eta_i\dvtx i=1,\ldots
,m)$ from the posterior distribution of $(\Lambda_{t_s},\Sigma,\eta)$
at the point $0\leq t_s \leq1$ and use them to estimate $\widehat
{E}_s(U)=\frac{1}{m}\sum{U(\Lambda_{t_s,i},\Sigma_i,\eta_i,y)},
\forall s=1,\ldots,S+1$.

\item[(ii)] \textit{Geometric mean path}: This path is constructed over
the distributional space (Gelman and Meng, \citeyear{r14}), hence, we model the density for
the model $M_t$ at each point along the grid. We use the density
$f_t(\Lambda,\Sigma, \eta|Y)=f(y|\Lambda,\Sigma,\eta)^t\pi(\Lambda,\Sigma
,\eta)$ as the unnormalized density for the model $M_t$ connecting the
prior and the posterior, when $\pi(\Lambda,\Sigma,\eta)$ and
$f(y|\Lambda,\break\Sigma,\eta)$ are the prior and the likelihood function,
respectively. By using PS along this path we can find the log marginal
for the models $M_0$ and $M_1$, as the normalizing constant for the
prior is known. Hence, the $\log{\mathit{BF}}$ can be estimated by using those
estimates of the log marginal for those models. The score function
$U(\Lambda, \Sigma, \eta,Y, t)$ will be the log likelihood function
$\log{f(y|\Lambda,\Sigma,\eta)}$.
\end{longlist}

The theorem below verifies the regularity conditions of path sampling
for factor models. For PS to succeed we also need convergence of MCMC
at each point in the path. That will be taken up after proving the theorem.

\begin{theorem}\label{th2.1}
Consider path sampling for factor models with parametric arithmetic
mean path (PAMP) and likelihood as given above for factor models.
Assume prior is proper and the corresponding score function is
integrable w.r.t. the prior:
\begin{longlist}[(1)]
\item[(1)] The interchangeability of integration and differentiation in (\ref
{eqn22}) is valid.
\item[(2)]$E_t(U)$ is finite as $t\rightarrow0$.\vadjust{\goodbreak}
\item[(3)] The path sampling integral for factor models, namely, (\ref
{eqn23}), is finite.
\end{longlist}
\end{theorem}

\begin{pf}
Here, for notational convenience, we\break write $(\Lambda,\Sigma, \eta
)=\theta$. When $f(Y|\theta)$ and $\pi(\theta)$ are the likelihood
function of the data and the prior density function for the
corresponding parameter, respectively, then the following is equivalent
to showing equation (\ref{eqn22}):
\begin{eqnarray*}
\frac{d}{dt}\int^{\infty}_{-\infty}f(Y|\theta,t)\pi(
\theta)\,d\theta=\int^{\infty}_{-\infty}\frac{d}{dt}f(Y|
\theta,t)\pi(\theta)\,d\theta.
\end{eqnarray*}
We can write the LHS as the following:
\begin{eqnarray*}
&=&\lim_{\delta\to0}\int^{\infty}_{-\infty}
\frac{f(Y|\theta,t+\delta
)-f(Y|\theta,t)}{\delta}\pi(\theta)\,d\theta
\\
&=&\lim_{\delta\to0} \int^{\infty}_{-\infty}
f'\bigl(Y|\theta,t'\bigr)\pi(\theta )\,d\theta\
,t'\in[t,t+\delta]
\\
&=&\lim_{\delta\to0} \int^{\infty}_{-\infty}U\bigl(Y|
\theta,t'\bigr)f\bigl(Y|\theta ,t'\bigr)\pi(\theta)\,d
\theta,
\end{eqnarray*}
where $t'\in[t,t+\delta]$. $U$ is a quadratic function in $\theta$ and,
hence, its absolute value is bounded above by a quadratic function in
$\theta$, free of $t$ but depending on~$Y$. $f(Y|\theta,t')$ is bounded
by the global maximum of the MVN likelihood, say, $M$, achieved at
$\widehat{\Omega}$ [equation (\ref{eq2.6})]. Now applying the moment assumptions
for $\pi(\theta)$, we can use the Dominated Convergence theorem (DCT)
and take the limit within the integral sign. The rest of statements $2$
and $3$ follow similarly.
\end{pf}

In Remark~\ref{rem1} we extend the theorem to other paths. Then in a series of
remarks we study various aspects like convergence and divergence of PS,
that are closely related to the theorem. All the remarks are related to
the theorem and insights gained from its proof. Remark~\ref{rem5} is the most important.

\begin{remark}\label{rem1} For PS with GMP, the score function is the log
likelihood function which can be\break bounded as before by using the RHS of
equation (\ref{eqn24}). Also, $f(y|\Lambda,\Sigma,\eta)^t\leq(1\vee
f(y|\hat{\Omega}))$ with $\hat{\Omega}$ as in equation (\ref{eqn24}).
We believe a similar generalization holds for most paths modeling means
of two models. Now the proof of Theorem~\ref{th2.1} applies exactly as before
(i.e., as for PAMP). We exhibit performance of PS for this path in
Section~\ref{sec4}.
\end{remark}

\begin{remark}\label{rem2} If we further assume the MCMC average at each point
on the grid converges to the Expectation of the score function of MCMC,
then the theorem implies the convergence of PS. We showed the integrand
is continuous on [0, 1].\vadjust{\goodbreak} So by continuity it can be approximated by a
finite sum. Now take the limit of the MCMC average at each of these
finitely many grid points. However, even if the MCMC converges in
theory, the rate of convergence may be very slow or there may be a
problem with mixing even for $m=50\mbox{,}000$, which we have taken as our
gold standard for good MCMC. This problem will be apparent to some
extent from high MCMC standard deviation.
\end{remark}

\begin{remark}\label{rem3} As $t\rightarrow0$ the likelihood is practically
independent of the extra parameters of the bigger model, so that a
prior for those parameters (conditional on other parameters) will not
learn much from data. In particular, the posterior for these parameters
will remain close to the diffuse prior one normally starts with. If the
prior fails to have the required finite moment in the theorem, the
posterior will also be likely to have large values for moments, which
may cause convergence problems for the MCMC. That's why we chose a
prior making the score function integrable. In the proof of the
theorem, we have assumed the first two moments of the prior to be
finite. In most numerical work our prior is a \textit{t} with 5 to 10 d.f.
\end{remark}

\begin{remark}\label{rem4} In the same vein, we suggest that even when the
integral at $t$ near zero converges, the convergence may be slow for
the following reason. Consider a fixed $(\Lambda_t,\Sigma,\eta)$ with a
large posterior or negative value of $U(\Lambda_t,\Sigma,\eta)L(\Lambda_t,\Sigma,\eta)$ at point $t$, the same large value will occur at
$(\frac{1}{t}\Lambda_t,\Sigma,\eta)$ with prior weight $\pi(\frac
{1}{t}\Lambda_t,\Sigma,\eta)$. For priors like $t$-distribution with low
degrees of freedom, $\pi(\frac{1}{t}\Lambda_t,\Sigma,\eta)$ will not
decay rapidly enough to substantially diminish the contribution of the
large value of $U(\Lambda_t,\Sigma,\eta)L(\Lambda_t,\break\Sigma,\eta)$ at
$(\Lambda_t,\Sigma,\eta)$.
\end{remark}

\begin{remark}\label{rem5} The structure of the likelihood and prior actually
provides insight as to when the MCMC will not converge to the right
distribution owing to bad mixing. To this end, we sketch a heuristic
argument below, which will be supported in Section~\ref{sec3.4} by MCMC figures:

\begin{longlist}[(1)]
\item[(1)] The maximized likelihood remains the same along the whole path,
because the path makes a one-to-one transformation of the parameter
space as given below.
\item[(2)] If the MLE of $\lambda_h$ at $t=1$ is $\hat{\lambda}_h$, then the
MLE at $t=t'$ is $\frac{\hat{\lambda}_h}{t'}$ (subject to variation due
to MCMC at two different points at the path), which goes to infinity as
$t$ goes to zero. This happens as the $\hat{\lambda}_h$ remains the
vector among $\lambda'_h$\vadjust{\goodbreak} (where $\lambda'_h$ is the MCMC sample from
model $M_t$ at $t$) having the highest maximum likelihood. Hence, as $t
\to0$, $\pi(\hat{\lambda}_h/t)\to0$ at a rate determined by the tail
of the prior. The conflict between prior and maximized likelihood may
also be viewed as a conflict between the nested models, with the prior
favoring the parsimonious smaller model. This inherent conflict in
model selection seems to have the following implications for MCMC.
\end{longlist}
\end{remark}

We expect to see a range (say, $[t_1,t_2]$) near zero showing a
conflict between prior and maximized likelihood. Definitely the points
$t_1$ and $t_2$ are not well specified, but we treat them as such so as
to under\-stand some underlying issues of mixing and convergence here. On
the set of points $t>t_2$ the MCMC samples are expected to be around
the points maximizing likelihood, whereas for $t<t_1$ they will be
nearly zero due to the concentration around a value $\lambda_h$ which
is both the prior mode and the mle under $M_0$, namely, $\lambda_h=0$.
But for any point in the range $[t_1,t_2]$, they will span a huge part
of the parameter space, ranging from points maximizing likelihood to
ones having higher prior probability, showing a lot of fluctuations
from MCMC to MCMC. The MCMC outputs in Section~\ref{sec3.4} show both
clusters but having highly fluctuating values (Figure~\ref{fig1}, Section~\ref{sec3.4}) for the proportions of the clusters.

Equation (\ref{eqn25}) tells us that the score function is proportional
to $\frac{\lambda'_h}{t}$ (where $\lambda'_h$ is the MCMC sample from
model $M_t$ at $t$). Hence, we will see $E_t(U)$ as an increasing
function while $t\to t_2$ from the right-hand side [(2) in Remark~\ref{rem5}].
This leads to a lot of variation of the estimate of $E_t(U)$ for
different MCMC samples in the range $[t_1,t_2]$ as explained above.
Also, as explained above, for $t<t_1$, the score function will
concentrate near zero.

The width of the zone of conflict (here $t_2-t_1$) will shrink, if we
have a relatively strong decaying tail of the prior. On the other hand,
for heavy-tailed priors we may see these above mentioned fluctuations
for a longer range, causing a loss of mass from the final integration.
These problems are aggravated by the high dimension of the problem and
the diffuse spread of the prior on the high-dimensional space. This may
mean the usual BF estimated by PS will be off by an order of magnitude.
We will see the implications reflected in some figures and tables in
the next section, when we study PAMP-PS for factor models in detail in
Section~\ref{sec3}.

\begin{remark}\label{rem6} We have checked that adaptive\break choice of grid points
by \citet{r24}, which improves\vadjust{\goodbreak} accuracy in their two examples with GMP,
does not help in the case of the very large fluctuations described
above. It seems to us that adaptive choice would work better when the
two models tested are less dissimilar than the models in Remark~\ref{rem5}, for
example, when the smaller of two nested models is true (Section~\ref{sec3.1})
or when our proposed modification of PS is used (Section~\ref{sec3.3}).
However, we have not verified this because even without adaptive
choice, our new proposal worked quite well in our examples.
\end{remark}

We note in passing that in both the examples of \citet{r24}, the two
models being tested have maximum likelihoods that differ by fifteen in
the log scale, whereas for the models in Remark~\ref{rem5} they differ by much
more, over a hundred.

\section{What Do Actual Computations Tell Us?}\label{sec3}

Following the discussion in the previous section, we would like to
study the effects of the theoretical observations in the previous
section for the implementation of path sampling. Here we only consider
the PAMP for PS, and for notational convenience we will mention it as
just PS. After studying estimated BF's in several simulated data sets
(not reported here) from various factor models, we note a few salient
features. Error in estimation of the BF or the discrepancy between
different methods tends to be relatively large, if one of the following
is true: the data has come from the complex model rather than the
simpler model, the prior is relatively diffuse or the value of the
precision parameters are relatively small. Different subsections study
what happens if the complex or simpler model is true, the effect of the
prior, the grid size and the MCMC size. These are done in Sections~\ref{sec3.1}--\ref{sec3.3}.

In Section~\ref{sec3.3} we introduce a new PS scheme, which operates through
a chain of paths, each path involving two nested models with a small
change between the contiguous pairs. The new scheme is denoted as Path
Sampling with Small Changes (PS-SC). The effect of precision parameters
will also be studied in this subsection for PS-SC. Then we study the
MCMC samples and try to understand their behavior from the point of
view of explaining the discrepancy between different methods for
estimating Bayes factors and why PS-SC does better than PS in
Section~\ref{sec3.4}.

Our simulated data are similar to those of \citet{r18} but have
different parameters. \textit{We use a \textup{2}-factor model and a \textup{1}-factor
model as our complex model $M_1$ and simpler model $M_0$, respectively,
to demonstrate the underlying issues.} The loading parameters and the
diagonal entries of the $\Sigma$ matrix are given in Tables~\ref{tab2} and~\ref{tab3}. In
simulation we take model $M_0$ or $M_1$ as true but $\Sigma$ is not
changed. Of course, if the one-factor model $M_0$ is true, then since
it is nested in $M_1$, $M_1$ is also true.

\begin{table}
\caption{Loading factors used for simulation}\label{tab2}
\begin{tabular*}{\columnwidth}{@{\extracolsep{\fill}}lccccccc@{}}
\hline
Factor 1 & 0.89 & 0\phantom{0.} & 0.25 & 0\phantom{0.} & 0.8 & 0\phantom{0.} & 0.5\\
Factor 2 & 0\phantom{0.0} & 0.9 & 0.25 & 0.4 & \phantom{0.}0 & 0.5 & 0\phantom{0.}\\
\hline
\end{tabular*}
\end{table}

\begin{table}
\caption{Diagonal entries of $\Sigma$}\label{tab3}
\begin{tabular*}{\columnwidth}{@{\extracolsep{\fill}}lcccccc@{}}
\hline
0.2079 & 0.19 & 0.15 & 0.2 & 0.36 & 0.1875 & 0.1875\\
\hline
\end{tabular*}
\end{table}

\subsection{Issues in Complex (2-Factor) Model}\label{sec3.1}
\begin{figure}

\includegraphics{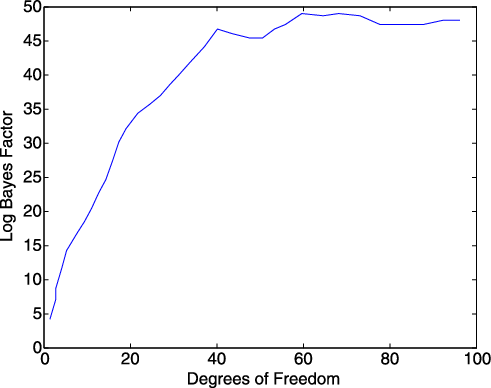}

\caption{Dependance of $\log\mathit{BF}_{21}$ over prior for 3rd data set.}
\label{fig1}
\end{figure}

We will study the effect of grid size, prior and the behavior of MCMC,
keeping in mind Theorem~\ref{th2.1} and the remarks in Section~\ref{sec2}. For path
sampling with the PAM path, we now discuss the effect of the prior and
the two tuning parameters, namely, the effect of the grid size and MCMC
size, on the estimated value of the BF and their standard deviation.
Following the discussion in Remarks~\ref{rem3} and~\ref{rem4}, we know that $\lim_{t\rightarrow0}E_t(U)$
is finite and path sampling converges under
some finite moment assumption for the prior. The prior considered in PS
by \citet{r17} are Cauchy and half-Cauchy, which do not have any finite
moments and so U is not integrable. We therefore choose a relatively
diffuse prior, but with enough finite moments for $U$. For finite mean
and variance one needs a $t$ with at least four degrees of freedom. Our
favorites are $t$-distributions with 5 to 10 degrees of freedom. We show
results for 5 and 10 d.f. only. But we first explore the sensitivity of
the estimate to changes in d.f. of the $t$-distribution as prior, over a
range of 1 through 90. The BF values change considerably until we reach
about 40 d.f. and then it stabilizes. In Table~\ref{tab4} we report the $\log
{\mathit{BF}}$ values estimated for 5 data sets simulated from a 2-factor model
using different priors. The behavior of the estimated $\log{\mathit{BF}}$ with
the change of d.f. continuously from 1 to 100 is shown in Figure~\ref{fig1} for
the 3rd data set.

\begin{table}
\caption{PAM-PS: Dependance of $\log\mathit{BF}_{21}$ over prior,\break 2-factor
model true}\label{tab4}
\begin{tabular*}{\columnwidth}{@{\extracolsep{\fill}}lcccc@{}}
\hline
\multicolumn{5}{c}{\textbf{PS using grid size 0.01}} \\
\hline
$\bolds{t_1}$ & $\bolds{t_5}$ & $\bolds{t_{10}}$ &$\bolds{t_{90}}$& \textbf{normal} \\
\hline
2.62& 14.42& 22.45&70.20 & 70.25\\
3.67& 11.90& 21.39&68.70 & 68.72\\
3.00& 13.43& 21.31&47.06 & 47.21\\
4.29& 13.17& 18.49&48.03 & 48.13\\
4.20& 13.11& 18.48&47.70 & 47.74\\
\hline
\end{tabular*}
\end{table}

We can see the estimate of the BF changing with the change in the
pattern of the tail of the prior. The effect of the grid size and MCMC
size on MCMC-standard deviation of the estimate are studied, using
priors $t_{10}$ and $N(0,1)$ and reported in Table~\ref{tab5}. We report the mean
of the estimates found from 25 different MCMC runs and the
corresponding standard deviation as MCMC-standard deviation. The study
has been done on the 2nd of the 5 data sets simulated from model 1 earlier.

\begin{table}[b]
\caption{PAM-PS: Dependence of $\log{\mathit{BF}_{21}}$ (MCMC-standard
deviation) estimates over grid size and MCMC size,\break 2-factor model true}\label{tab5}
\begin{tabular*}{\columnwidth}{@{\extracolsep{4in minus 4in}}lccc@{}}
\hline
\multicolumn{2}{@{}l}{\textbf{Grid size}} & \textbf{0.01} & \textbf{0.001}\\[-6pt]
\multicolumn{2}{@{}l}{\rule{99pt}{1pt}}&\multicolumn{1}{c}{\rule{44pt}{1pt}}&\multicolumn{1}{c@{}}{\rule{44pt}{1pt}}\\
\textbf{MCMC size} & \textbf{Prior} & \textbf{Data 2} & \multicolumn{1}{c@{}}{\textbf{Data 2}} \\
\hline
\phantom{0.}5000& $t_{10}$ & 21.26 (1.39) &21.26 (1.29) \\
& $N(0,1)$ & 66.89 (4.15) & 67.21 (3.28)\\ [5pt]
50,000& $t_{10}$ &23.71 (1.21)& 23.57 (0.52)\\
& $N(0,1)$ & 68.21 (3.62) & 68.23 (3.11) \\
\hline
\end{tabular*}\vspace*{-3pt}
\end{table}

As expected, Table~\ref{tab5} shows a major increase of MCMC size and finer grid
size reduces the MCMC-standard deviation of the estimator. The
difference between the mean values of BF estimated by $t_{10}$ and
$N(0,1)$ differ by an order of magnitude. We will study these issues as
well as special patterns exhibiting MCMC in Section~\ref{sec3.4}. Though the
different variants of PS compared here differ in their estimated value
of BF, they still choose the correct model 100\% of the time.

\subsection{Issues in Simpler (1-Factor) Model}\label{sec3.2}

Now we study the scenario when the 1-factor model is true focusing on
the effect of prior, grid size and MCMC size on the estimated Bayes
factor (Table~\ref{tab6}). In this scenario the estimates do not change much with
the change of prior, so we will report the estimates for prior $t_{10}$
and $N(0,1)$ with different values of MCMC size and grid size.

\begin{table}
\caption{PAM-PS: Dependence of $\log{\mathit{BF}_{21}}$ (MCMC-standard
deviation) estimates over grid size and MCMC size, while 1-factor model true}\label{tab6}
\begin{tabular*}{\columnwidth}{@{\extracolsep{\fill}}lccc@{}}
\hline
\multicolumn{2}{@{}l}{\textbf{Grid size}} & \textbf{0.01} &
\textbf{0.001}\\[-6pt]
\multicolumn{2}{@{}l}{\rule{95pt}{1pt}}&\multicolumn{1}{c}{\rule{52pt}{1pt}}&\multicolumn{1}{c@{}}{\rule{52pt}{1pt}}\\
\textbf{MCMC size} & \textbf{Prior} & \textbf{Data 1} & \textbf{Data 1} \\
\hline
\phantom{0.}5000& $t_{10}$ & $-4.26$ (0.054) & $-4.27$ (0.044) \\
& $N(0,1)$ & $-4.62$ (0.052) & $-4.60$ (0.051)\\[3pt]
50,000& $t_{10}$ &$-4.24$ (0.012)& $-4.24$ (0.007)\\
& $N(0,1)$ & $-4.60$ (0.006) & $-4.62$ (0.005) \\
\hline
\end{tabular*}
\end{table}

This table shows us that the MCMC-standard deviation improves with the
finer grid size and large MCMC size as expected, but the estimated
values of $\mathit{BF}_{21}$ remain mostly the same. As noted earlier, PS
chooses the correct model 100\% of the time when $M_0$ is true.

We explain tentatively why the calculation of BF is relatively stable
when the lower dim model $M_0$ is true. Since $M_0$ is nested in $M_1$,
$M_1$ is also true in this case, which in turn implies both max
likelihoods (under $M_0$ and $M_1$) are similar and smaller than for
data coming from $M_1$ true (but not $M_0$). This tends to reduce or at
least is associated with the reduction of the conflict between the two
models or prior and likelihood along the path mentioned in Remark~\ref{rem5}.

Moreover, the score function for small $t$ causes less problem since
for data under $M_0$, $\lambda'_2$ is relatively small compared with
that for data generated under~$M_1$.

So we see when two models are close in some sense, we expect their
likelihood ratio will not fluctuate widely provided the parameters from
the two parameter spaces are properly aligned, for example, if found by
minimizing a K-L divergence between the corresponding densities or
taking a simple projection from the bigger space to the smaller space.
This is likely to make importance sampling more stable than if the two
models were very different. It seems plausible that this stability or
its lack in the calculation of BF will also show up in methods like PS
that are derived from importance sampling in some way. Ingenious
modifications of importance sampling seems to mitigate but not
completely solve the problem. Following this idea of closer models in
some sense, we modify PS in a similar manner below.

\subsection{Path Sampling with Small Changes: Proposed Solution}\label{sec3.3}

In Remark~\ref{rem5}, Section~\ref{sec3.1}, a prior-likelihood conflict was identified
as a cause of poor mixing. This will be re-examined in the next
subsection. In the present subsection we propose a modification of PS
which tries to solve or at least reduce the magnitude of this problem.

\begin{table*}
\caption{$\log{\mathit{BF}_{21}}$ (MCMC-standard deviation) estimated by
PAM-PS-SC and PAM-PS }\label{tab7}\vspace*{-2pt}
\begin{tabular*}{\textwidth}{@{\extracolsep{\fill}}lcd{2.10}d{2.9}d{2.10}@{}}
\hline
\textbf{True model} & \multicolumn{1}{c}{\textbf{MCMC size}} & \multicolumn{1}{c}{\textbf{PS-SC}} &
\multicolumn{1}{c}{\textbf{PS} $\bolds{(t_{10})}$} & \multicolumn{1}{c@{}}{\textbf{PS} $\bolds{(N(0,1))}$} \\
\hline
1-factor & \phantom{0.}5000 & -8.09\ (0.013) & -4.26\ (0.054) & -4.62\ (0.052)\\
1-factor & 50,000& -8.08\ (0.0067)&-4.24\ (0.012)& -4.60\ (0.0065)\\
2-factor & \phantom{0.}5000 & 80.14\ (0.66) & 21.26\ (1.39) & 66.89\ (4.15)\\
2-factor & 50,000& 80.75\ (0.54)& 23.71\ (1.21)& 68.21\ (3.62)\\
\hline
\end{tabular*} \vspace*{-3pt}
\end{table*}

To solve this problem without having to give up our diffuse prior (we
will be using $t$ with 10 d.f. as our prior), we try to reduce the
problem to a series of one-dimensional problems so that the competing
models are close to each other. We calculate the Bayes factor by using
the path sampling step for every single parameter that may be zero,
keeping others fixed. It is easily seen that the original log Bayes
factor is the sum of all the log Bayes factors estimated in these
smaller steps. We denote this procedure as PS-SC (Path Sampling with
Small Change) and implement with the parametric arithmetic mean path
(PAMP). (As pointed out by a Referee, there is scope for exploring
other paths, including a search for an optimal one, to reduce the
MCMC-variance.) More formally, if we consider {$\lambda_2$} as a
$p$-dimensional vector, then $M_0$ and $M_1$ differ only in the last
$p-1$ parameters, as $\lambda_{21}$ is always zero due to the
upper-triangular condition. We consider $p$ models $M'_i\dvtx  i=1,\ldots,p$,
where for model $M'_i$ we have first $i$ parameters of $\lambda_2$
being zero correspondingly. If we define $\mathit{BF}'_{i,i+1}=\frac
{m_{i}(x)}{m_{i+1}(x)}$, when $m_{i}(x)$ is the marginal for the model
$M'_i$, then
\[
\log{\mathit{BF}_{21}}=\sum_{i=1}^{p-1}
\log{\mathit{BF}'_{i,i+1}}.
\]
So we perform $p-1$ path sampling computations to estimate $\log
{\mathit{BF}'_{i,i+1}},\forall i=1,\ldots, p-1$. And for each of the steps the
score function will be of the following form:
\begin{eqnarray*}
&& U_i'(\Lambda, \Sigma, \eta,Y, t)\\
&&\quad=\sum
^n_{j=1}(y_j-\Lambda_t
\eta_i)'\\
&&\qquad{}\cdot \Sigma^{-1}\bigl(0^{p\times(h-1)},[0_{i};
\lambda_{h,i+1};0_{p-i-1}]\bigr)\eta_i,
\end{eqnarray*}
where $\Lambda_t=(\lambda_1,[0_{i};t\lambda_{2,i+1};\lambda_{2,(i+2,\ldots,p)}])$.

As in the case of the small model true, the max likelihoods under both
models are close, and generally the two models are close, suggesting
fluctuations are less likely and true BF is not very large. This seems
generally to lead to stability of computation of BF.

Also, the parameter $\lambda'_2$ is now one dimensional. So the score
function is more likely to be small than when $\lambda'_2$ is a vector
as under PS. We also notice that in each step the score function is not
anymore proportional to $\frac{\lambda'_2}{t}$ but rather to $\frac
{\lambda'_{2i}}{t}$ which will be much smaller in value, hence reducing
the fluctuation and loss of mass.

Computational implementation shows it to be stable for different MCMC
size and grid size regarding MCMC-standard deviation and also produces
a smooth curve of $E_t(U)$ for every single step. Here we use an MCMC
size of 5000$/$50,000 and grid size of 0.01 for our study and report the
corresponding estimated BF values for two data sets from 1-factor and
2-factor models, respectively. The MCMC-standard deviation of the
estimates along with the mean of the estimated value over 25 MCMC runs
are reported in Table~\ref{tab7}. PS-SC has smaller standard deviation than PS
under both $M_0$ and $M_1$. In Section~\ref{sec2} and Section~\ref{sec3.4}, we argue
that, at least under $M_1$, PS-SC provides a better estimate of BF.

Now we see the effect of changing the precision parameters keeping the
factor loadings as before. The diagonal entries of $\Sigma$ are in
Table~\ref{tab8}.\vadjust{\goodbreak} The precision of these 3 models lie in the ranges of [2.77,
6.55], [1.79, 2.44], [1.36, 1.66], respectively.

\begin{table}
\tabcolsep=0pt
\caption{Diagonal entries of $\Sigma$ in the 3 different models: the
first one is modified from Ghosh and Dunson
(\protect\citeyear{r17})}\label{tab8}\vspace*{-2pt}
\begin{tabular*}{\columnwidth}{@{\extracolsep{\fill}}ld{1.4}ccd{1.2}d{1.3}d{1.4}d{1.4}@{}}
\hline
Model 1 & 0.2079 & 0.19 & 0.15 & 0.2 & 0.36 & 0.1875 & 0.1875\\
Model 2 & 0.553 & 0.52 & 0.48 & 0.54 & 0.409 & 0.55 & 0.54\\
Model 3 & 0.73 & 0.71 & 0.67 & 0.7 & 0.599 & 0.67 & 0.72\\
\hline
\end{tabular*}    \vspace*{-3pt}
\end{table}

We study PS-SC for 6 data sets generated from the 3 models (2 data sets
with $n=100$ from each model: Data 1 from 1-factor and Data 2 from
2-factor model) and report the estimated Bayes factor value in Table~\ref{tab9}.

\begin{table*}
\caption{$\log\mathit{BF}_{21}$ (MCMC-standard deviation) estimation by PS-SC:
effect of precision parameter}\label{tab9}
\begin{tabular*}{\textwidth}{@{\extracolsep{\fill}}lccd{3.11}d{1.10}@{}}
\hline
\multicolumn{1}{@{}l}{\textbf{Model}} & \textbf{True model}& \textbf{Data} & \multicolumn{1}{c}{\textbf{PS-SC}} &
\multicolumn{1}{c@{}}{\textbf{PS} $\bolds{(t_{10})}$} \\
\hline
Model 1& 1-factor &Data 1 & -8.09\ (0.012) & -3.84\ (0.055) \\
& 2-factor &Data 2 & 71.59\ (0.66) & 19.81\ (1.38) \\[3pt]
Model 2& 1-factor &Data 1 & -11.01\ (0.0066) & -3.09\ (0.0277)\\
& 2-factor &Data 2 & 51.41\ (0.3658) & 2.8\ (1.9104)\\[3pt]
Model 3& 1-factor &Data 1 & -5.13\ (0.0153) & -2.6\ (0.0419)\\
& 2-factor &Data 2 & 3.975\ (0.0130) & 2.2\ (0.3588)\\
\hline
\end{tabular*}      \vspace*{-3pt}
\end{table*}

The effect of precision parameters is seen on the estimated value of
the Bayes Factor (BF), more prominently when the 2-factor model is
true. Generally, the absolute value of the BF decreases with the
decrease in the value of the precision parameters. For the smaller
value of precision parameters, we expect the model selection to be less
conclusive, explaining the pattern shown in the estimated BF values.

Under $M_1$, PS is often bad in estimating the Bayes Factor
($\mathit{BF}_{21}$), but since the true Bayes factor is large, it usually
chooses the true model as often as PS-SC. When $M_0$ is true, PS is
much better in estimating the Bayes factor, but since the Bayes factor
is usually not that large, it does not choose $M_0$ all the time. The
probability of choosing $M_0$ correctly depends on the data in addition
to the true values of the parameters. PS-SC does better than PS in all
these cases; it estimates $\mathit{BF}_{21}$ better and chooses the correct
model equally or more often. The sense in which PS-SC estimates
$\mathit{BF}_{21}$ better has been discussed in detail earlier in this section.
Under $M_0$ PS-SC estimates $\mathit{BF}_{21}$ better by having a smaller, that
is, more negative, value than PS.

\subsection{Issues Regarding MCMC Sampling}\label{sec3.4}

\begin{figure*}[b]\vspace*{-3pt}

\includegraphics{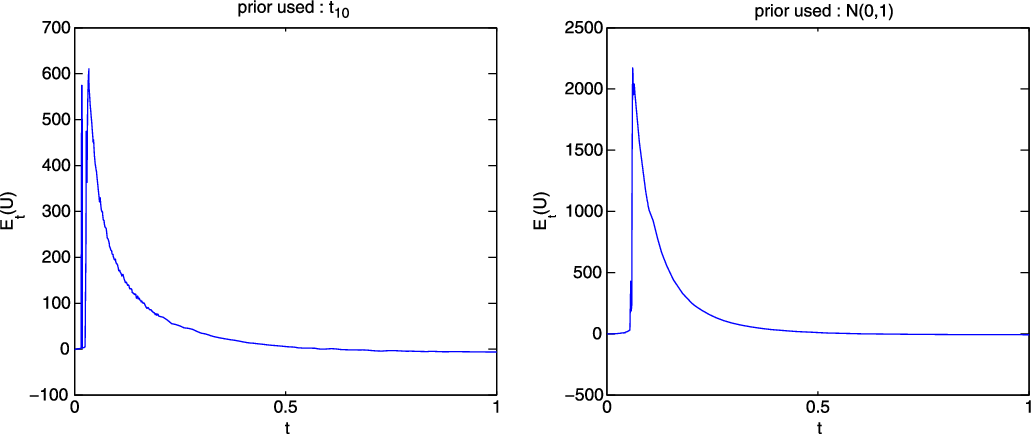}

\caption{$E_t(U)$ for prior $t_{10}$ and $N(0,1)$, 2-factor model is true.}\label{fig2}
\end{figure*}

\begin{figure*}

\includegraphics{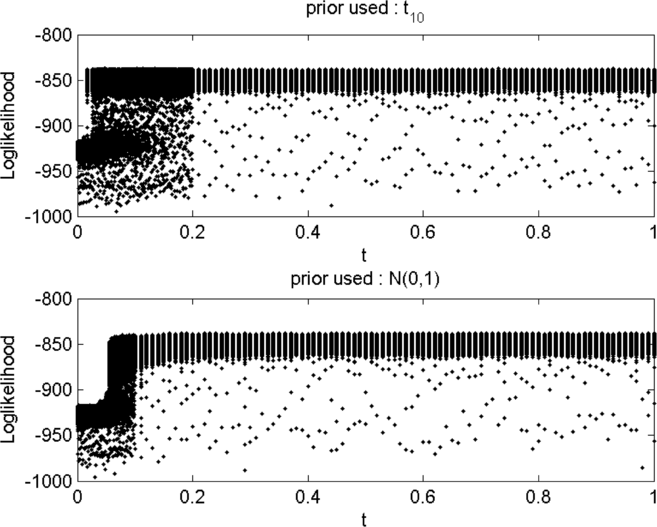}

\caption{$\mathit{Log}\mbox{-}\mathit{likelihood}$ for prior $t_{10}$ and $N(0,1)$, 2-factor
model is true.}\label{fig3}\vspace*{-3pt}
\end{figure*}

This subsection is best read along with the remarks in Section~\ref{sec2}. We
first study the graph of $E_t(U)$ and the likelhood values for the MCMC\vadjust{\goodbreak}
samples at $t$ for both the $t_{10}$ and $N(0,1)$ prior (Figures~\ref{fig2} and~\ref{fig3}).
We will plot the likelihood as a scalar proxy since we can not show
fluctuations of the vector of factor loadings in the MCMC output. The
clusters of the latter can be inferred from the clusters of the former.
\textit{We will argue that there are two clusters at each grid point
and the mixing proportion of the two clusters has a definite pattern.}

\begin{figure*}[b]

\includegraphics{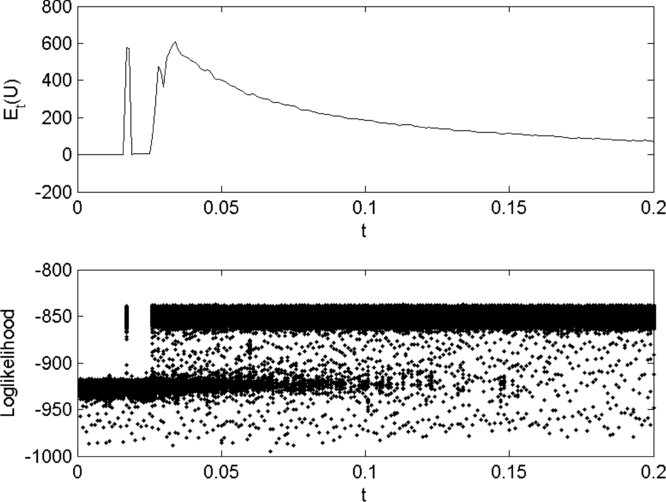}

\caption{$E_t(U)$ and $\mathit{Log}\mbox{-}\mathit{likelihood}$ for prior $t_{10}$ in the range $t\in[0,0.2]$, 2-factor model is true.}\label{fig4}
\end{figure*}

Under the true 2-factor model $M_1$, denote $\lambda'=[\lambda'_1,\lambda'_2]$, where $\lambda'_i$ is the loading for the
corresponding latent factor under $M_t$. Here $\lambda'_2$ is a 7$\times
$1 vector and becomes zero, as it approaches $M_0$ from $M_1$ (as $t\to
0$). The posterior distribution at each $M_t$ can be viewed roughly as
a sort of mixture model with two components representing $M_0$ and
$M_1$, the form of the likelihood as given in Theorem~\ref{th2.1}. In the diagram
(Figure~\ref{fig4}) of the log-likelihood of MCMC samples, we see two clear
clusters around log-likelihood values $-850$ and $-925$, representing MCMC
outputs with nonzero $\lambda'_2$ and zero\vadjust{\goodbreak} $\lambda'_2$ values,
respectively. We may think of them as coming from the component
corresponding to $M_1$ (cluster 2) and the component corresponding to
$M_0$ (cluster 1). Samples of both clusters are present in the range
[0.03, 0.2], while samples appear to be predominantly from cluster 2 until
$t=0.1$. A good representation of samples from cluster 1 are only
present in the range [0, 0.1]. In the range [0.03, 0.2], both clusters occur
with proportions varying a lot. Moreover, here the magnitude of the
score function is proportional to $\frac{\lambda'_2}{t}$. We see these
fluctuations in Figure~\ref{fig4} in the region [0.03, 0.2]. This is also brought
out by the MCMC standard deviation of $E_t(U)$ \textit{which are of
order of \textup{30--50} in the log scale}.

We notice the absence of any samples from $M_1$ for $t<0.03$, except
some chaotic representation for a few random values of $t$ (notice in
the figure, a spike representing samples from $M_1$ at $t=0.016$),
clearly representing poor mixing of MCMC samples near the model~$M_0$.\vadjust{\goodbreak}

\begin{figure*}

\includegraphics{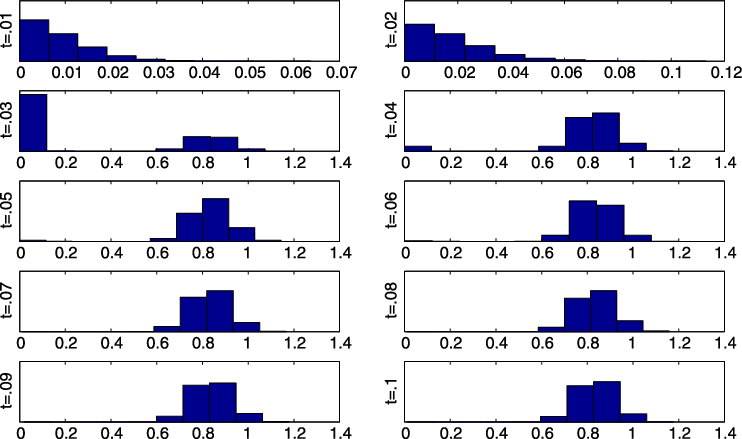}

\caption{Histograms for $\lambda'_{22}$ for different values of $t$
near $t=0$ (MCMC size used 50,000), using PS.}\label{fig5}
\end{figure*}

\begin{figure*}[b]

\includegraphics{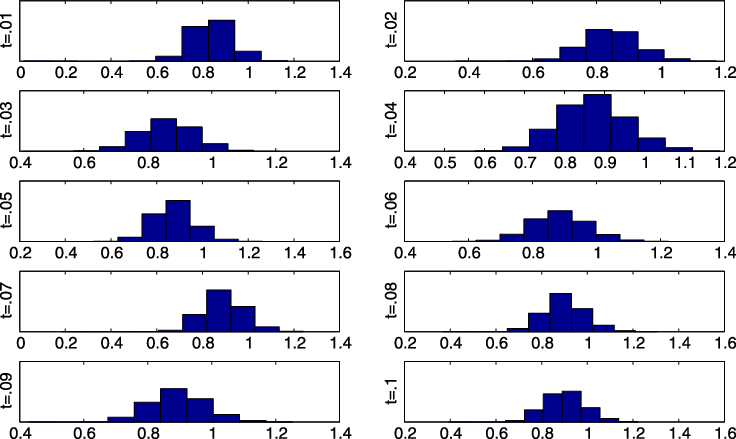}

\caption{Histograms for $\lambda'_{22}$ for different values of $t$
near $t=0$ (MCMC size used 50,000), using PS-SC.}\label{fig6}
\end{figure*}

The new method PS-SC stabilizes the estimated Bayes factor value with a
very small MCMC-stand\-ard deviation. Here we check through Figures~\ref{fig5}
and~\ref{fig6} that it avoids prior-likelihood conflict and the problem about mixing
for MCMC samples seen for the standard PS. We concentrate our study for
the first step of PS-SC. In this step only the first component of
$\lambda'_2$, $\lambda'_{22}$ converges to zero as $t \to0$. So here
we consider the spread of the MCMC sample of $\lambda'_{22}$ for
different values of $t$ near $t=0$, from both PS and PS-SC in Figures~\ref{fig5}
and~\ref{fig6} by considering the histogram of MCMC sample of $\lambda'_{22}$. We
can easily notice that the spread of the MCMC sample fluctuates in
between the two modes in a chaotic manner showing poor or unstable
mixing for PS, whereas PS-SC samples come from both the clusters and
slowly shift toward the prior mode as $t \to0$. We have also studied
but do not report similar nice behavior regarding mixing of MCMC of
PS-SC for the data simulated from the 1-factor model.

\begin{figure*}

\includegraphics{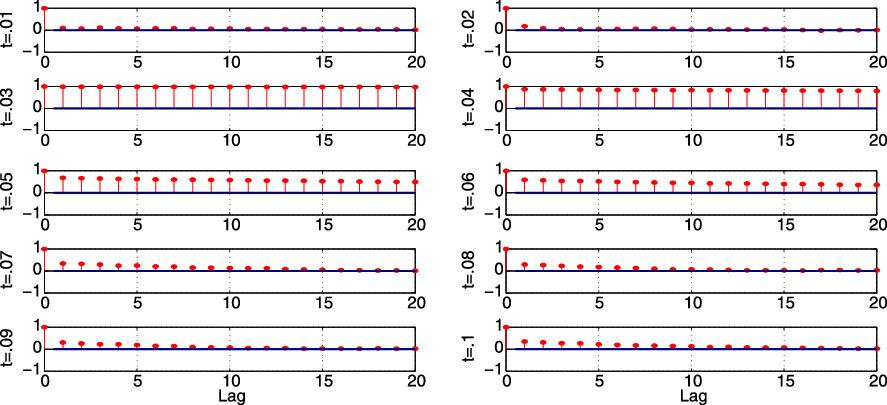}

\caption{Autocorrelation for $\lambda'_{22}$ for different values of
$t$ near $t=0$ (MCMC size used 50,000), using PS.}\label{fig7}
\end{figure*}

\begin{figure*}[b]

\includegraphics{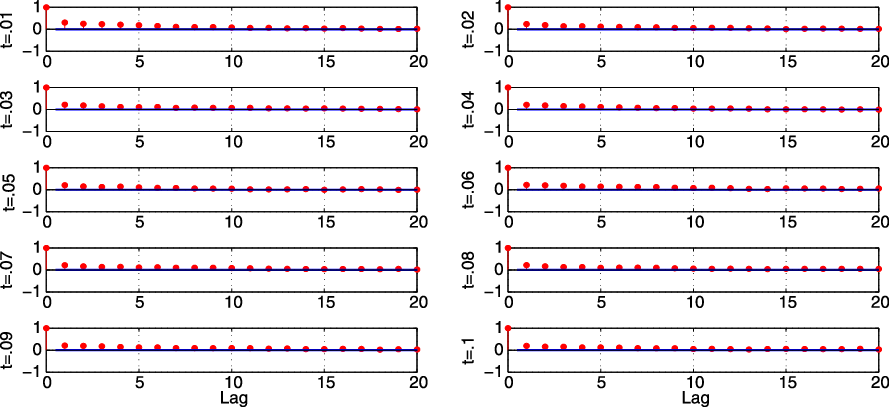}

\caption{Autocorrelation for $\lambda'_{22}$ for different values of
$t$ near $t=0$ (MCMC size used 50,000), using PS-SC.}\label{fig8}
\end{figure*}

The poor mixing discussed above for MCMC outputs for PS will now be
illustrated with plots of autocorrelation for $\lambda'_{22}$ for
different lags (Figure~\ref{fig7}). For the sake of comparison, we do the same
for PS-SC (Figure~\ref{fig8}). Clearly, except very near $t=0$, that is, in what
we have called the chaotic zone, the autocorrelations for PS are much
bigger than those for PS-SC. However, near $t=0$, though plots in both
Figures~\ref{fig7} and~\ref{fig8} are small, those for PS are slightly smaller. We have no
simple explanation for this.\looseness=1

Poor mixing seems to lead to missing mass and random fluctuations for
calculations for $E_t(U)$. This probably explains the discrepancy we
have noticed in the estimation of BF by PS as compared with PS-SC. We
now look at autocorrelations for a first factor loading in Figure~\ref{fig7} and
second factor loading in Figure~\ref{fig8}. The top rows in each of the two
figures show zero autocorrelation, as they are very close to $t=0$. On
the other hand, high autocorrelations are shown in the next two rows.
We believe they correspond to what we called a chaotic region. The
bottom two rows of Figure~\ref{fig8} show small autocorrelation. They correspond
to the second factor loading which comes only in model 2, and they also
depict the zone dominated by model 2. The other figure is in the same
zone as in the previous line, but the variable considered is a 1-factor
loading. Here autocorrelation also eventually tends to 0, but its
values are bigger than in Figure~\ref{fig8}. We do not have any simple
explanation for this higher autocorrelation.

The above discussion covers the case when the more complex model is
true. If the simpler model ($M_0$) is true, as noted in Section~\ref{sec3.3}
both PS and PS-SC perform well in estimating the Bayes factor as well
as choosing the correct model. The Bayes factor based on PS-SC provides
stronger support for the true model than the Bayes factor based on PS.

To check whether PSSC works well in other examples as in the factor
model, we try to explore its impact on our earlier toy example. In this
case, we were unable to implement path sampling with small changes, but
rather used a pseudo-PSSC scheme. Going back to our example where we
have taken $m=7$ and $p=10$, we define a sequence of models as the following:
\begin{eqnarray}
M_i\dvtx y_i\sim N \biggl(0,\Sigma= \pmatrix{
A_{11} & 0 \vspace*{2pt}
\cr
0 & A_{22} } \biggr)\nonumber\\
\eqntext{\mbox{when }
A_{11} \mbox{ is }(i\times i)\mbox{ matrix for }i=7,8,9,10.}
\end{eqnarray}
We can see our previously defined $M_0$ and $M_1$ are now $M_7$ and
$M_{10}$, respectively.
For our pseudo-PSSC, we estimate $\log{\mathit{BF}}_{i,i+1}$ by $\log{\mathit{BF}}$ between
the models $M'_0$ and $M'_1$, with $m=i$ and $p=i+1$:
\begin{eqnarray*}
M'_t\dvtx y_i\sim N \biggl(0,\Sigma=
\pmatrix{ A_{11} & tA_{12} \vspace*{2pt}
\cr
t(A_{12})' & A_{22} } \biggr).
\end{eqnarray*}
Still being underestimates on each step, this method improves on the
standard path sampling in terms of Bayes factor estimation, as we can
see in the Table~\ref{tab10}.

%
\begin{table}
\tabcolsep=4pt
\caption{Performance of PS and pseudo-PS-SC in toy example modeling
covariance: Log Bayes factor\break (MCMC standard deviation)}\label{tab10}
\begin{tabular*}{\columnwidth}{@{\extracolsep{\fill}}ld{3.9}d{3.9}@{}}
\hline
\multicolumn{1}{@{}l}{\textbf{Method}} & \multicolumn{1}{c}{\textbf{Data 1}} & \multicolumn{1}{c@{}}{\textbf{Data 2}} \\
 \hline
True BF value& 258.38& -132.87 \\
PS estimate of BF& 184.59\ (0.012)& -20.11\ (0.008)\\
pseudo-PSSC estimate of BF& 195.35\ (0.011) & -25.21\ (0.007)\\
\hline
\end{tabular*}
\end{table}

\section{Implementation of Other Methods}\label{sec4}

We have explored several methods of estimating the ratio of
normalizing constants, for example, the methods of \citet{r31}, \citet
{r10}, \citet{r34} and \citet{r8}. The method of \citet{r34} models a link
function of means, but here we are concerned with models for the
variance--covariance matrix. We could not use Chib's method here since
for our parameter expanded prior the full conditionals of the original
model parameters are not available. But we were able to implement the
deterministic variational Bayes method of \citet{r31} and the Laplace
approximation with a correction due to \citet{r10}. Since the results
were not satisfactory, we do not report them in this paper. In the
variational Bayes approach, the method selected the correct model
approximately 80\% of the time, but the estimated logBF values were
considerably over (or under) estimated. The variational Bayes method is
worth further study, possibly with suitable modifications. It appears
to us it is still not understood when Belief Propagation provides a
good approximation to a marginal or not, for example, \citet{r13}
commented: \textit{Only recently we have witnessed an explosion of
research for theoretical understanding of the performance of the BP
algorithm in the context of various combinatorial optimization
problems, both tractable and intractable (NP-hard) versions}.

Following the discussion in Section~\ref{sec2.4}, we have implemented the
GMP-PS. Here the marginal for both models is estimated by constructing
a path between the prior distribution to the posterior distribution of
the model. Due to very high-dimensionality of the model, the modes of
prior and posterior distribution are far apart. So as discussed before,
the MCMC sampling along the path fails to sample\break smoothly and
fluctuates between the two modes in a chaotic way near the prior mode.
Hence, the estimate of the marginal of both the models becomes very
unstable. Due to the poor estimation of BF, this method also fails to
choose the correct model very often. As in the case of GMP-PS, the AIS
with the GM path also did not work well. Hence, we implemented the AIS
with the PAM-path. Implementation of PAM-AIS is also very time
intensive, so we have only implemented PAMP-AIS with MCMC sample size
5000. PAM-AIS not only shows very high MCMC-standard deviation, but it
also fails to choose the correct model many a time, when the 2-factor
model is correct. The last methods we looked at are the following:

\begin{longlist}[(1)]
\item[(1)] Importance Sampling (IS).
\item[(2)] Newton-Raftery approximation (BICM).
\item[(3)] Laplace/BIC type approximation (BICIM).
\end{longlist}

IS is the most easy to implement and shows moderately good results in
choosing the correct model \citep{r17}). We study the stability of
Bayes factor values estimated by IS with the change of the MCMC size in
Table~\ref{tab11}.

Similarly, we also study the stability of the estimates of the Bayes
factor by BICM and BICIM (explained in~\ref{appa1.3} in the Appendix) using
MCMC sample size 10,000, where both of these methods show significantly
less amount of MCMC-standard deviation than other methods considered.
Hence, we will only consider PS-SC, BICM and BICIM to explore model
selection for a dimension much higher than previously considered.

\begin{table}
\caption{Study of IS, BICM and BICIM for different MCMC size: Estimated
Bayes factor (MCMC standard deviation)}\label{tab11}
\begin{tabular*}{\columnwidth}{@{\extracolsep{\fill}}ld{3.10}d{2.12}@{}}
\hline
\textbf{Method(MCMC-size)/}&&\\
\textbf{true model} & \multicolumn{1}{c}{\textbf{2-factor model}} & \multicolumn{1}{c@{}}{\textbf{1-factor model}} \\
\hline
IS (10,000)& 109.78\ (168.72) & 0.0749\ (0.1063) \\
IS (50,000)& 97.12\ (61.25) & -5.39\ (84.35) \\
IS (100,000)& 86.92\ (110.35) & -3.07\ (10.41) \\
IS (200,000)& 83.66\ (58.53) & -2.69\ (2.96) \\
BICM (10,000) & 68.66\ (0.93) & -5.72\ (0.62) \\
BICIM (10,000) & 67.9\ (0.11)& -5.3\ (0.57) \\
PS-SC (5000) & 80.75\ (0.63)& -8.08\ (0.0013) \\
\hline
\end{tabular*}
\end{table}

%
\begin{table*}[b]
\caption{Simulated model ($p=20$, $n=100$) and ($k={}$the number of true
factors): Comparison of log Bayes factor}\label{tab12}
\begin{tabular*}{\textwidth}{@{\extracolsep{\fill}}lcd{3.10}d{3.2}d{3.2}@{}}
\hline
\textbf{Data} & \textbf{BF} & \multicolumn{1}{c}{\textbf{PS-SC}} & \multicolumn{1}{c}{\textbf{BICM}} & \multicolumn{1}{c@{}}{\textbf{BICIM}} \\
\hline
Data1 ($k=1$)& $\mathit{BF}_{21}$ & -25.91\ (0.0233)& -32.68& -38.01\\
& $\mathit{BF}_{32}$ & -24.84\ (0.0594)& -21.18& -38.24\\
& $\mathit{BF}_{43}$ & -22.79\ (0.0483)& -19.81& -43.77 \\[3pt]
Data2 ($k=2$)& $\mathit{BF}_{21}$ & 225.81\ (4.2099)& 248.09& 219.87 \\
& $\mathit{BF}_{32}$ & -23.61\ (0.0160)& -23.59& -46.17 \\
& $\mathit{BF}_{43}$ & -19.18\ (0.0297)& -20.3& -47.98 \\[3pt]
Data3 ($k=3$)& $\mathit{BF}_{21}$ & 152.07\ (1.7422)& 185.45& 162.3 \\
& $\mathit{BF}_{32}$ & 104.17\ (2.5468)& 198.1& 168.54 \\
& $\mathit{BF}_{43}$ & -17.35\ (0.0276)& -29.73& -48.24 \\
\hline
\end{tabular*}
\end{table*}

\section{Effect of Precision Parameters and High-Dimensional (Simulated
and Real) Data Set}\label{sec5}

Our goal is to explore if PS-SC may be made more efficient by combining
with BICM and BICIM and also to explore the number of dimensions much
higher than before and the real life examples.

In the examples in this section, $p$ varies from 6 to 26. We have 2
examples of real life examples with $p=6$ and 26 and a simulated
example with $p=20$. As expected, PS-SC still takes a long time, even
with a parallel processing for high-dimensional examples. We explore
whether PS-SC can be combined with BICM and BICIM to substantially
reduce time, since their performance seems much faster than PS-SC.

We compare the behavior of these methods for a higher-dimensional model
and for some real data sets taken from \citet{r18} and \citet{r1}. We
first consider one 3-factor model with $p=20$ and $n=100$ in Table~\ref{tab12}.

We notice that all the methods are selecting correct models for all the
3 data sets, but based on our earlier discussion of PS-SC, we believe
only this method provides a reliable estimate of BF. Now we will
compare the methods for some real data sets. We choose two data sets:
``Rodent Organ Data'' from \citet{r18} and ``26-variable Psychological
Data'' from \citet{r1}. These data sets have been normalized first before
analyzing them further. We not only study the estimated Bayes factor
but also the model chosen by them.

In the ``Rodant Organ Data'' the model chosen by PS-SC and other
methods are, respectively, the 3-factor model and 2-factor model (Table~\ref{tab13}). For
the ``26-variable Psychological Data,'' PS-SC and BICM/BICIM choose the
model with 3 factors and 4 factors, respectively (Table~\ref{tab14}). The models chosen by
PS-SC and the other methods are close, but as expected differ a lot in
their estimate of BFs.

There is still no rigorously proved Laplace approximation for
relatively high-dimensional cases because of analytical difficulties.
Problems of determining sample size in hierarchical modeling, pointed
out by Clyde and George (\citeyear{r9}), are avoided by both versions of our
approximations (Appendix~\ref{appa1.3}). These two methods seem to be good as a
preliminary searching method to narrow the field of plausible models
before using PS-SC. This saves time relative to PS-SC for model search
as seen in the previous examples.

\section{Conclusion}\label{sec6}

We have studied PS for factor models (and one other toy example) and
have identified the component of PS that is most likely to go wrong and
where. This is partly based on the fact that we have a relatively
simple sufficient condition for factor models (Theorem~\ref{th2.1}). Typically,
for the higher-dimensional model the MCMC output for finding the
integral along grid points in the path may become quite unreliable at
some parts of the path. Some insight about why it happens and how it
can be rectified has been suggested. MCMC seems to be unreliable for PS
when the higher-dimensional model is true. The problem is worse the
more the two models differ, as when a very high-dimensional model is
being compared to a low-dimensional model.

%
\begin{table}
\caption{Rodant organ weight data ($p=6$, $n=60$):\break Comparison of log
Bayes factor}\label{tab13}
\begin{tabular*}{\columnwidth}{@{\extracolsep{\fill}}lccc@{}}
\hline
\textbf{Bayes factor} &\textbf{PS-SC}&\textbf{BICM} & \multicolumn{1}{c@{}}{\textbf{BICIM}} \\
\hline
$\log \mathit{BF}_{21}$ & 4.8& 26.34 & 21.57 \\
$\log \mathit{BF}_{32}$ & 10.52& $-3.14$ & $-10.01$ \\
$\log \mathit{BF}_{43}$ & $-3.28$& & \\
\hline
\end{tabular*}
\end{table}
%
\begin{table}
\caption{26-variable psychological data ($p=26$, $n=300$):\break Comparison
of log Bayes factor}\label{tab14}
\begin{tabular*}{\columnwidth}{@{\extracolsep{\fill}}ld{3.2}d{3.2}d{3.2}@{}}
\hline
\multicolumn{1}{@{}l}{\textbf{Bayes factor}} &\multicolumn{1}{c}{\textbf{PS-SC}}&\multicolumn{1}{c}{\textbf{BICM}} & \multicolumn{1}{c@{}}{\textbf{BICIM}} \\
\hline
$\log \mathit{BF}_{21}$&122.82 &205.27 & 188.19 \\
$\log \mathit{BF}_{32}$&35.27& 71.05 & 35.5 \\
$\log \mathit{BF}_{43}$&-10.7& 23.16 & 7.55 \\
$\log \mathit{BF}_{54}$&-33.32& -4.63 & -25.51 \\
$\log \mathit{BF}_{65}$&-16.7& -17.32& -43.21 \\
\hline
\end{tabular*}
\end{table}

The suggestion for rectification was based on the intuition that PS,
like importance sampling itself, seems more reliable when the two
marginal densities in the Bayes factor are relatively similar, as is
the case when the smaller of two nested models is true. Based on this
intuition, we suggested PS-SC and justified PS-SC by comparing MCMC
output and MCMC standard deviation of both PS-SC and PS.\looseness=1

It is our belief that the above insights as to when things will tend to
go wrong and when not, will also be valid for the other general
strategy for selection from among nested models, namely, RJMCMC. Piyas
Chakraborty in Purdue is working on a change point problem in several
parameters where \citet{r35} have an accurate approximation to the Bayes
factor, which may be used for validation. He will explore small changes
as well as adaptive MCMC.

Our work has focused on model selection by Bayes factors, which seems
very natural since it provides posterior probability for each model.
However, model selection is a complicated business and one of its major
purposes is also to find a model that fits the data well. Several model
selecting statisticians feel this should also be done along with
calculation of Bayes factors.

However, there has not been a good discussion on how one should put
together the findings from the two different approaches. We hope to
return to these issues in a future communication.

\textit{A natural future direction of our study of factor models is to
add to the model an unknown mean vector with a regression setup. The
problem now would be to simultaneously determine a parsimonious model
for both the variance--covariance matrix and the mean vector. There are
natural priors for these problems, but computation of the Bayes factor
seems to be a challenging problem.}

\begin{appendix}

\section*{Appendix}\label{app}

\subsection{Other Methods}\label{appa1}
\subsubsection{Importance sampling}\label{appa1.1}
Suppose we have two densities proportional to two functions $f(x)$ and
$g(x)$, which are feasible to evaluate at every $x$, but one of the
distributions, say, the one induced by $f(x)$, is not easy to sample.
Then the importance sampling (IS) estimate of the ratio of normalizing
constants is based on m independent draws $x_1,\ldots,x_m$ generated
from the distribution defined by $g(x)$. We first compute the importance
weights $w_i=\frac{f(x_i)}{g(x_i)}$ and then define the IS estimate:
%
\begin{equation}
\frac{1}{m}\sum_{i=1}^m
w_i.
\end{equation}
Under the assumption that $g(x)\neq0$ when $f(x)\neq0$, $\frac
{1}{m}\sum_{i=1}^m w_i$ converges as $m\rightarrow\infty$ to $Z_f/Z_g$,
when $Z_f=\int{f(x)\,dx}$ and $Z_g=\int{g(x)\,dx}$ are the normalizing
constants for $f(x)$ and $g(x)$. The variability of the IS estimate
depends heavily on the variability of the weight functions. So to have
a good IS estimate, we need to have $g(x)$ as a good approximation to
$f(x)$, which is difficult to achieve in problems with high or
moderately high-dimensional, possibly multimodal density.

Analysis of Bayesian factor models using IS has been introduced by \citet
{r17}. The IS estimator of BF for factor models is based on m samples
$\theta_i^{(h)}$ from the posterior distribution, under $M^{(h)}$
%
\begin{equation}
\widehat{\mathit{BF}}_{h-1,h}=\frac{1}{m}\sum^m_{i=1}
\frac{p(y|\theta_i^{(h)},k=h-1)}{p(y|\theta_i^{(h)},k=h)},
\end{equation}
which in turn is based on the following identity:
\begin{eqnarray}
&&\int\frac{p(y|\theta^{(h)},k=h-1)}{p(y|\theta^{(h)},k=h)}p\bigl(\theta^{(h)}|y,k=h\bigr)\,d
\theta^{(h)}\nonumber
\\
& &\quad =\int p\bigl(y|\theta^{(h)},k=h-1\bigr)\frac{p(\theta^{(h)})}{p(y|k=h)}
\,d\theta^{(h)}
\\
\nonumber
& & \quad=\frac{p(y|k=h-1)}{p(y|k=h)}.
\nonumber
\end{eqnarray}

\citet{r17} implemented IS with a parameter expanded prior. They also
have noted that IS is fast and often (90\%) chooses the correct model
in simulation. In our simulation IS chooses a true bigger model
correctly, but a 20\% error rate was observed when the smaller model is true.

\subsubsection{Annealed importance sampling}\label{appa1.2}

Following\break Neal (\citeyear{r30}), we consider densities $p_t\dvtx t\in[0,1]$ joining the
densities $p_0$ and $p_1$. We choose densities by discretizing the path
$p_{t_{(i)}}$ where $0=t_{(1)}<\cdots<t_{(k)}=1$ and then simulate a
Markov chain designed to converge to $p_{t_{(k)}}$. Starting from the
final states of the previous simulation, we simulate some number of
iterations of a Markov chain designed to converge to $p_{t_{(k-1)}}$.
Similarly, we simulate some iterations starting from the final steps of
$p_{t_{(j)}}$ designed to converge to $p_{t_{(j-1)}}$ until we simulate
some iterations converging to $p_{t_{(1)}}$. This sampling scheme
produces a sample of points $x_1,\ldots, x_m$ and then we compute the
weights $w_i=\frac{p_1(x_i)}{p_0(x_i)}$. Then the estimate of the ratio
of normalizing constant becomes as follows:\looseness=1
%
\begin{equation}
\frac{1}{m}\sum_{i=1}^m
w_i.
\end{equation}
Notice that while both AIS and PS are based on MCMC runs along a path
from one model to another, the MCMC'S are drawn at each point, but the
details are very different. Due to the better spread of MCMC samples,
the estimates in AIS seem to be better than those calculated by IS when
the smaller model is true, helping in correct model selection and also
improving the estimation of Bayes factors. However, simulations show
that AIS has the same problem as IS in estimating the Bayes factor when
the bigger model is true.

\subsubsection{BIC type methods: Raftery--Newton and our method using
information matrix}\label{appa1.3}

In contrast to the methods previously discussed, we try to directly
estimate the marginal under each model and then use these marginals to
find the Bayes factor. We know that BIC is an approximation to the
log-marginal based on a Laplace-type approximation of the log-marginal
\citep{r19}, under the assumption of i.i.d. observations. Thus,
%
\begin{eqnarray}\label{eqa.5}
\log\bigl(m(x)\bigr)&\approx &\log\bigl(f(x|\hat{\theta})\pi(\hat{\theta})
\bigr)
\nonumber
\\
&&{}+(p/2)\log(2\pi )+(p/2)\log(n)\\
&&{}+\log\bigl(\bigl|H^{-1}_{1,\hat{\theta}}\bigr|^{1/2}
\bigr),\nonumber
\end{eqnarray}
where $H_{1,\hat{\theta}}$ is the observed Fisher Information matrix
evaluated at the maximum likelihood estimator using a single
observation. For BIC we just use
%
\begin{eqnarray}\qquad
\log\bigl(m(x)\bigr)&\approx& \log\bigl(f(x|\hat{\theta})\pi(\hat{\theta })
\bigr)+(p/2)\log(n)
\nonumber
\\[-8pt]
\\[-8pt]
\nonumber
& \approx& \log\bigl(f(x|\hat{\theta})\bigr)+(p/2)\log(n),
\end{eqnarray}
ignoring other terms as they are $O(1)$.

It is known BIC may be a poor approximation to the log-marginal in
high-dimensions (Berger, Ghosh and Mukhopadhyay, \citeyear{r5}). To take care of this problem, \citet{r32}
suggest the following. Simulate i.i.d. MCMC samples from the posterior
distributions, evaluate independent sequence of $\log(\mathrm{prior} \times
\mathrm{likelihood})$s (log-p.l.) $\{l_t\dvtx t=1,\ldots,m\}$, and then an estimate
for the marginal is
%
\begin{equation}\label{eqa.7}
\log\bigl(m(x)\bigr)\approx\bar{l}-s^2_l\bigl(\log(n)-1
\bigr),
\end{equation}
where $\bar{l}$ and $s^2_l$ will be the sample mean and variance of
$l_t$'s. We call this method BICM, following the convention of \citet{r32}.

In order to apply (\ref{eqa.5}), we do not need to evaluate $n$ since it cancels by
combining the last two terms. This suggests the approximation (\ref{eqa.5}) take
care of the point raised by \citet{r9}. However, (\ref{eqa.7}) does use $n$, but we
do not know the impact on the approximation.

We have also used the Laplace approximation (\ref{eqa.5}) without any change as
likely to work better than the usual BIC. We compute the Information
Matrix at the maximum prior${}\times{}$likelihood (mpl) value under the
model and impute its value in the computation of the marginal. To find
the mpl estimate, we use the MCMC sample from the posterior
distribution and pick the maxima in that sample. Then we search for the
mple in its neighborhood, using it as the starting point for the
optimization algorithm. In our simulation study, it has been seen to
give very good results similar to the computationally intensive
numerical algorithms used to find the maximum of a function over the
whole parameter space seen by taking repetition of MCMC runs and large
MCMC samples. In the spirit of \citet{r32}, we call this method BICIM,
indicating the use of Information Matrix based Laplace Approximation.
We also used several other modifications that did not give good
results, so are not reported.

\subsection{A Theoretical Remark on the Likelihood Function}\label{appa2}
It appears that the behavior of the likelihood, for example, its
maximum, plays an important role in model selection, specifically in
the kind of conflict we see between PS and the Laplace approximations
(BICM, BICIM) when the bigger model is true (and the prior is a $t$ with
a relatively small d.f.). The behavior seems to be different from the
asymptotic behavior of maximum likelihood under the following standard
assumptions. Assume dimension of the parameter space is fixed and usual
regularity conditions hold. Moreover, when the big model is true but
the small model is assumed (so that it is a misspecified model), the
Kullback--Liebler projection of the true parameter space to the
parameter space of the small model exists (Bunke and Milhaud, \citeyear{r6}).\looseness=1

\begin{fact*} Assume the big model is true, and the small model is
false. Then, as may be verified easily by the Taylor expansion,
\begin{longlist}[(1)]

\item[(1)] $\log L(\hat{\theta}_{\mathrm{big}})-\log L(\theta_{\mathrm{true}(\mathrm{big})})=O_P(1)$

\item[(2)] $\log L(\hat{\theta}_{\mathrm{small}})-\log L(\mbox{KL projection of
}\break
\theta_{\mathrm{true}(\mathrm{big})} \mbox{to }
\Theta_{\mathrm{small}})=O_P(1)$

\item[(3)] $\log L(\theta_{\mathrm{true}(\mathrm{big})})-\log L(\mbox{KL projection of}\break
\theta_{\mathrm{true}(\mathrm{big})}\mbox{ to }\Theta_{\mathrm{small}})=O_P(n)$

and
\item[(4)]
\begin{eqnarray}
\nonumber
&&\log L(\hat{\theta}_{\mathrm{big}})-\log L(\hat{
\theta}_{\mathrm{small}})
\nonumber\\
& &\quad =\log L({\theta}_{\mathrm{true}(\mathrm{big})})\nonumber\\
&&{}\qquad-\log L(\mbox{KL projection of }{\theta
}_{\mathrm{true}(\mathrm{big})}\ \mathrm{to}\ \Theta_{\mathrm{small}})\\
&&\qquad{}+O_P(1)
\nonumber\\
& &\quad =O_P(n).
\nonumber
\end{eqnarray}
\end{longlist}

The maximized likelihood for factor models substantially overestimates
the true likelihood, unlike relation (1) above. Unfortunately, as
pointed out in \citet{r11}, the asymptotics of mle for factor models is
still not fully worked out.
\end{fact*}
\subsection{Matrix Used for the Toy Example}\label{appa3}
\fontsize{10}{12}{\selectfont{
\begin{eqnarray*}
\Sigma^{0}= \lleft(\matrix{ 128.35& 52.69& -19.25& -11.86& 24.34
\vspace*{2pt}
\cr
52.69& 73.37& -21.04& -37.85& 12.29\vspace*{2pt}
\cr
-19.25&
-21.04& 30.86& 8.63& -1.41\vspace*{2pt}
\cr
-11.86& -37.85& 8.63& 80.49& 4.66
\vspace*{2pt}
\cr
24.34& 12.29& -1.41& 4.66& 15.45\vspace*{2pt}
\cr
8.80& 8.74&
-13.58& 3.26& 2.58\vspace*{2pt}
\cr
10.63& 15.60& -3.03& -49.24& 2.05\vspace*{2pt}
\cr
13.75& 12.09& -11.64& -9.68& 3.72\vspace*{2pt}
\cr
-7.40& -14.08& 21.28& 22.18&
-1.31\vspace*{2pt}
\cr
-29.80& -17.27& 22.05& 8.52& -7.87& }\rright. %
\\[4pt]
\lleft.\matrix{8.80& 10.63& 13.75& -7.40& -29.80\vspace*{2pt}
\cr
8.74& 15.60&
12.09& -14.08&-17.27\vspace*{2pt}
\cr
-13.58& -3.03& -11.64& 21.28& 22.05
\vspace*{2pt}
\cr
3.26& -49.24& -9.68& 22.18& 8.52\vspace*{2pt}
\cr
2.58& 2.05&
3.72& -1.31& -7.87 \vspace*{2pt}
\cr
31.37& 11.62& -4.85& -16.89& -20.10
\vspace*{2pt}
\cr
11.62& 58.09& 7.00& -19.58&5.16\vspace*{2pt}
\cr
-4.85& 7.00&
26.59& -3.04& 11.17\vspace*{2pt}
\cr
-16.89& -19.58& -3.04& 31.81& 22.86
\vspace*{2pt}
\cr
-20.10& 5.16& 11.17& 22.86& 64.68 } \rright)
\end{eqnarray*}}}

\subsection{Choice of Prior Under $M_0$}\label{appa4}
A referee has asked whether under $M_0$, the prior for the extra
parameter can be chosen in a same optimal or philosophically compelling
manner. This has been a long-standing problem, but the method followed
for factor models is one of the standard procedures, apparently first
suggested by \citet{r11a}.

This prior is mentioned by \citet{r11a} and may be justified as follows.
One tries to ensure the extra parameter has similar roles under both
the models. If the joint prior of $(\theta_1, \theta_2)$ under $M_1$ is
$\pi(\theta_1, \theta_2)$, then the natural prior for $(\theta_2|\theta_1)$
is the usual conditional density of $\pi(\theta_2|\theta_1)$. In
our case $\pi(\theta_1, \theta_2)=\pi(\theta_1)\pi(\theta_2)$. So $\pi
(\theta_2|\theta_1)$ is as we have chosen. This is one of the standard
default choices. Another default choice is due to Jeffreys (\citeyear{r22}), but
when $\theta_1, \theta_2$ are independent, both lead to the same
choice. If we introduce a prior (e.g., minimizing MCMC-variance), it
may not be acceptable to Bayesian philosophy.
\end{appendix}

\section*{Acknowledgments}We thank Joyee Ghosh for helping us in
discussions on factor models and sharing her code and Andrew
Lewandowski for thought-provoking comments on an earlier draft.\vadjust{\goodbreak}

%


\end{document}